\documentclass[11pt]{article}

\usepackage{cite}
\usepackage{epsf}
\usepackage{amsmath}
\usepackage{amsfonts}
\usepackage{amssymb}

\usepackage{graphicx}
\usepackage{epsfig}
\usepackage{latexsym}

\def\corresponds{{\lower.2ex\hbox{=}}{\rm\kern-.75em^\triangle}}
\def\succsim{\succ\kern-.9em_\sim\kern.3em}
\def\precsim{\prec\kern-1em_\sim\kern.3em}
\def\slantfrac#1#2{\kern1em^{#1}\kern-.3em/\kern-.1em_{#2}}
\def\lfrac#1#2{{}^{#1\!}\kern-.0em/_{#2}}

\def\buildrel#1\under#2{\mathrel{\mathop{\kern0pt #2}\limits_{#1}}}

\usepackage[dvips,
   bookmarksopen=true,
   bookmarksnumbered=true,
   pdftitle={Extrapolation of the $Z\alpha$--Expansion},
   pdfauthor={Ulrich Jentschura},
   pdfstartview={FitB},
   colorlinks,
   linkcolor=blue]{hyperref}

\begin{document}

\bibliographystyle{myprsty}

\vspace*{1.0cm}
\begin{center}
\begin{tabular}{c}
\hline
\rule[-5mm]{0mm}{15mm}
{\Large \sf Extrapolation of the $Z\alpha$--Expansion}\\
{\Large \sf and Two--Loop Bound--State Energy Shifts}\\[2ex]
\hline
\end{tabular}
\end{center}
\vspace{0.2cm}
\begin{center}
Ulrich D. Jentschura
\end{center}
\vspace{0.2cm}
\begin{center}
{\it Theoretische Quantendynamik, \\
Fakult\"{a}t Mathematik und Physik der Universit\"{a}t Freiburg,\\
Hermann--Herder--Stra\ss{}e 3, 79104 Freiburg, Germany}
\end{center}
\vspace{0.3cm}
\begin{center}
\begin{minipage}{11.8cm}
{\underline{Abstract}}
Quantum electrodynamic (QED) effects that shift the binding energies  
of hydrogenic energy levels have been expressed in terms of a 
semi-analytic expansion in powers of $Z\alpha$ and $\ln[(Z\alpha)^{-2}]$,
where $Z$ is the nuclear charge number and $\alpha$ is the fine-structure
constant. For many QED effects, numerical data are available in the domain
of high $Z$ where the $Z\alpha$ expansion fails. In this Letter, we 
demonstrate that it is possible, within certain limits of accuracy, 
to extrapolate the $Z\alpha$-expansion from the low-$Z$ to the high-$Z$ domain.
We also review two-loop self-energy effects and provide an estimate for
the problematic nonlogarithmic coefficient $B_{60}$.
\end{minipage}
\end{center}
\vspace{1.3cm}

\noindent
{\underline{PACS numbers:}} PACS: 12.20.Ds, 11.15.Bt, 11.10.Jj\newline
{\underline{Keywords:}}
QED Radiative Corrections, \\
General properties of perturbation theory,\\
Asymptotic problems and properties.\\

\newpage

%
% Introduction 
%
\section{Introduction}
\label{introduction}

The subject of the current Letter is the investigation 
of QED radiative corrections in bound hydrogenlike systems  which
provide one of the most stringent and accurate available tests of quantum 
field theory and are amenable to high-precision spectroscopy 
on which the determination of fundamental constants is based~\cite{MoTa2000}.
The purpose of this investigation is twofold: first, to demonstrate that the 
$Z\alpha$-expansion which is inherently valid only at small $Z$,
can be extrapolated by ``deferred'' Pad\'{e} approximants
to the domain of high $Z$, albeit with a certain loss of accuracy
in the theoretical predictions. The second purpose is to provide a
brief review of logarithmic two-loop higher-order binding corrections
to the Lamb shift of hydrogenic states. The calculation of these
effects has recently been completed~\cite{Pa2001,Je2003jpa}, but
results have been provided only for the total effect which is the sum 
of the two-loop self-energy, two-loop vacuum polarization and combined
effects. For a comparison to numerical calculations which are
currently being pursued~\cite{YeInSh2003}, 
it is helpful to analyze the coefficients that relate
to specific sets of gauge-invariant diagrams.
This second purpose is actually a prerequisite for carrying out the 
extrapolation of the two-loop self-energy from low to high
$Z$, wherefore we commence this Letter with the endeavour of providing
some ``mini-review'' of the two-loop hydrogenic energy shifts.

%
% Brief Review of Two--Loop Hydrogenic Energy Shifts
%
\section{Brief Review of Two--Loop Hydrogenic Energy Shifts}
\label{review}

We use natural
Gaussian units with $\hbar = c = \epsilon_0 = 1$ and $e^2 = 4\pi\alpha$,
as it is customary for QED bound-state calculations.
The two-loop radiative shift of a hydrogenic S state, within the
$Z\alpha$-expansion, reads
\begin{eqnarray}
\label{DefESE2L}
\Delta E^{\rm (2L)}_{\mathrm{SE}} &=& \left(\frac{\alpha}{\pi}\right)^2 \,
(Z\alpha)^4 \, \frac{m}{n^3} \, H(Z\alpha)\,,
\end{eqnarray}
where
\begin{eqnarray}
\label{defH}
H(Z\alpha) &=& B_{40} +
(Z\alpha)\, B_{50} + (Z\alpha)^2 \,
\left\{ B_{63} \, \ln^3(Z\alpha)^{-2} \right. 
\nonumber\\[0.5ex]
& & \left. + B_{62} \, \ln^2(Z\alpha)^{-2}
+ B_{61} \, \ln(Z\alpha)^{-2} + B_{60} \right\} + \dots \,,
\end{eqnarray}
and the ellipsis denotes higher-order terms.

%${\mathcal R}(Z\alpha)$ is a remainder term that 
%vanishes as $Z\alpha \to 0$, i.e.
%%
%\begin{equation}
%{\mathcal R}(Z\alpha) \to 0 \quad \mbox{as} \quad 
%Z\alpha \to 0\,.
%\end{equation}
%%

\begin{figure}[htb!]
\begin{center}
\begin{minipage}{12.0cm}
\begin{center}
\centerline{\mbox{\epsfysize=15cm\epsffile{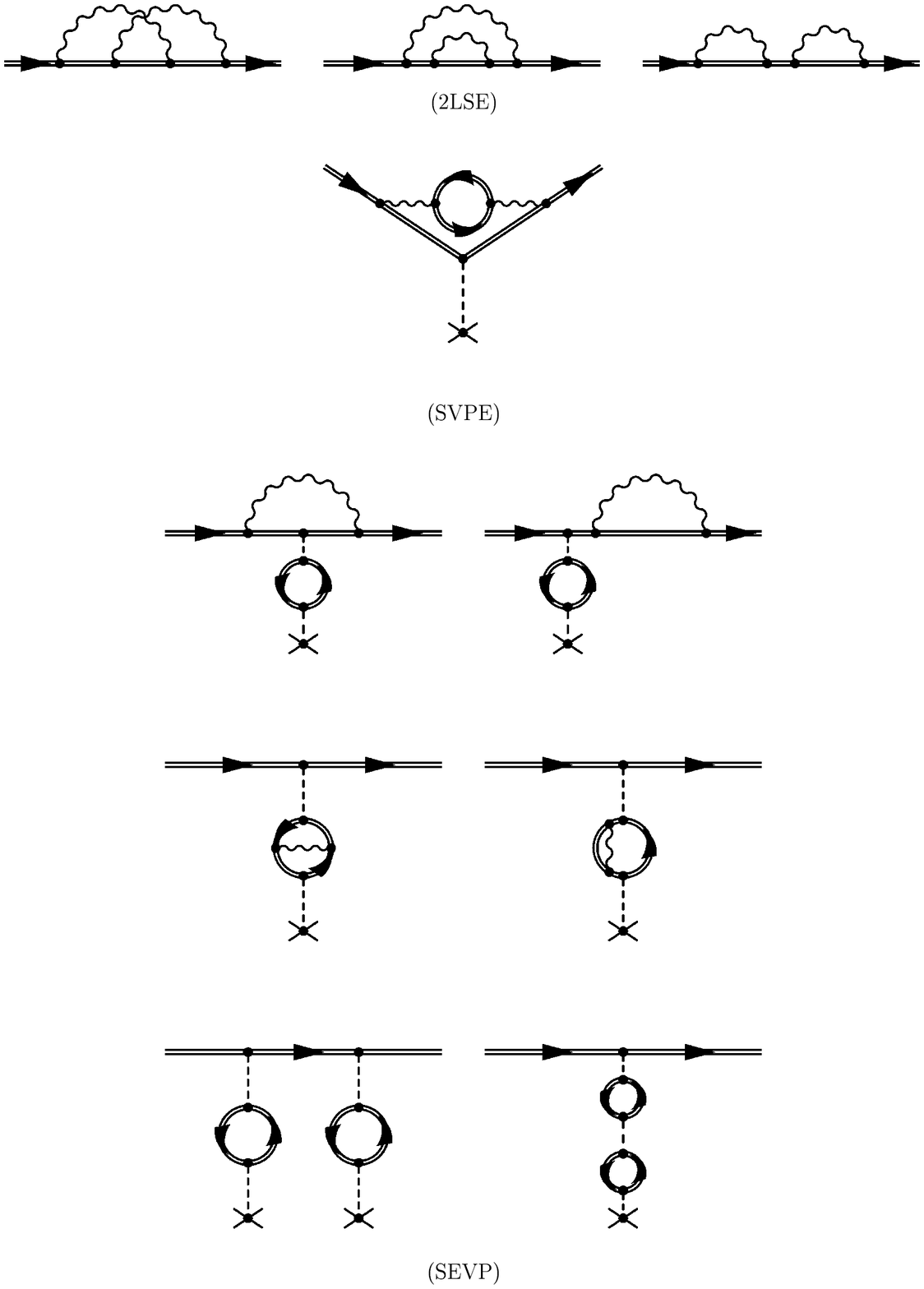}}}
\caption{\label{fig1}
The Feynman diagrams that contribute to the two-loop QED
energy shifts of hydrogenic bound states fall quite naturally 
into three separately gauge invariant categories: 
(i) the two-loop self-energy effects (2LSE), which are historically 
the most problematic, (ii) the vacuum-polarization insertion into 
the virtual photon line of the one-loop self-energy (SVPE),
and (iii) diagrams involving both the self-energy and the one-loop
vacuum polarization on the one hand and pure two-loop 
vacuum-polarization corrections on the other hand, 
summarized here as the set (SEVP).
The double line denotes the bound-state electron propagator,
i.e.~including all Coulomb interactions.}
\end{center}
\end{minipage}
\end{center}
\end{figure}

This section is a brief review of the known two-loop coefficients
$B_{40}$, $B_{50}$, $B_{63}$, $B_{62}$ and $B_{61}$.
The Feynman diagrams which contribute to the two-loop bound-state energy shifts
are shown in Fig.~\ref{fig1}.

%
% figure 1: OM decay
%

For S states, the contributions to $B_{40}$ can be evaluated
by considering the form-factor approach described e.g.~in 
Sec.~VIII.B.1--VIII.B.3 of~\cite{EiGrSh2001} or 
in Sec.~1~of~\cite{JePa2002}.
Detailed information about the $F'_1(0)$ form factor slope and the 
magnetic form factor $F_2(0)$ attributable to the 
different sets of diagrams (2LSE) and (SVPE) shown in 
Fig.~\ref{fig1} are given in Eqs.~(16),
(17) and (18) of Ref.~\cite{JePa2002}.

We recall the following $n$-independent results for 
the $B_{40}$-coefficients of S states,
\begin{subequations}
\label{B40}
\begin{eqnarray}
B^{\rm (2LSE)}_{40}(n{\rm S}) &=&
- \frac{163}{72} - \frac{85}{36}\,\zeta(2) +
9\,\ln(2)\,\zeta(2) - \frac94 \, \zeta(3) =
1.409\,244\,, \\[0.5ex]
B^{\rm (SVPE)}_{40}(n{\rm S}) &=&
- \frac{7}{81} + \frac{5}{36} \, \zeta(2) =
0.142\,043\,, \\[0.5ex]
B^{\rm (SEVP)}_{40}(n{\rm S}) &=&
- \frac{82}{81} =
-1.012\,346\,, \\[0.5ex]
B_{40}(n{\rm S}) &=&
- \frac{2179}{648} - \frac{20}{9}\,\zeta(2) +
9\,\ln(2)\,\zeta(2) - \frac94 \, \zeta(3) =
0.538\,941\,.
\end{eqnarray}
\end{subequations}
Note that the distribution of $B_{40}$-contributions 
among the different sets of diagrams in Fig.~\ref{fig1} 
[notably (2LSE) and (SVPE)]
is different from the separation into a ``self-energy''
correction [Eq.~(A24) of~\cite{MoTa2000}] and a ``magnetic moment
contribution'' [Eq.~(A25) of~\cite{MoTa2000}].
We also recall that the first treatment of the leading 
two-loop self-energy coefficient $B_{40}$ was completed in~\cite{ApBr1970}.

The evaluation of the relativistic correction $B_{50}$ 
due to the two-loop self-energy has represented a 
considerable challenge~\cite{Pa1994prl,EiSh1995}.
Diagrams involving a closed fermion loop were studied 
in~\cite{Pa1993pra,EiGrSh1997}. It might be useful to point out
that the contribution of the diagram (SVPE) is the 
sum of the contribution labeled $E_{\rm II}$ and 
$E_{\rm V}$ in Ref.~\cite{Pa1993pra}. The coefficients read: 
\begin{subequations}
\label{B50}
\begin{eqnarray}
B^{\rm (2LSE)}_{50}(n{\rm S}) &=& - 24.2668(31)\,, \\[0.5ex]
B^{\rm (SVPE)}_{50}(n{\rm S}) &=& -0.1571\,, \\[0.5ex]
B^{\rm (SEVP)}_{50}(n{\rm S}) &=& 2.8677\,, \\[0.5ex]
B_{50}(n{\rm S}) &=& -21.5562(31)\,.
\end{eqnarray}
\end{subequations}
These results are in agreement with the data presented in Eqs.~(A28)---(A29)
of~\cite{MoTa2000}.

The coefficients of sixth order in $Z\alpha$ have recently been 
analyzed in~\cite{Pa2001} (see also the references therein). 
The triple logarithm $B_{63}$ originates exclusively from the 
set (2LSE). The diagrams of the set (SEVP) have been 
calculated in Sec.~VII of~\cite{Pa2001}, and the results for 
the double and single logarithms originating from these diagrams 
can be obtained by adding the contributions labeled 
$E^1_{VP}$ and $E^2_{VP}$ in Ref.~\cite{Pa2001},
and the additional logarithm implicitly contained in the 
Eq.~(40) {\em ibid.} which is proportional to 
$B^{\rm (SEVP)}_{40}$. The diagram (SVPE) generates only 
a single logarithm given by $B^{\rm (SVPE)}_{40}/2$ [again consider
Eq.~(40) of Ref.~\cite{Pa2001}]. In total, the results read as follows:
\begin{equation}
\label{B63}
B_{63}(n{\rm S}) = B^{\rm (2LSE)}_{63}(n{\rm S}) = -\frac{8}{27} = -0.296296\,.
\end{equation}
The total result for $B_{62}$ as well as its $n$-dependence were
obtained in Refs.~\cite{Pa2001,Ka1996,Ka1996b,Ka1997}.
Here, we give the formulas for the particular sets of diagrams,
for the case $n=1$ as well as the difference to a state of general $n$:
\begin{subequations}
\label{B62}
\begin{eqnarray}
B^{\rm (2LSE)}_{62}(1{\rm S}) &=& \frac{16}{27} - \frac{16}{9}\, \ln(2) 
= -0.639\,669\,, \\[0.5ex]
B^{\rm (2LSE)}_{62}(n{\rm S}) &=& B^{\rm (2LSE)}_{62}(1{\rm S}) 
\! + \! \frac{16}{9} \left( \frac34 + \frac{1}{4 n^2} -
\frac1n - \ln(n) \! + \! \Psi(n) + C\right), \\[0.5ex]
B^{\rm (SVPE)}_{62}(n{\rm S}) &=& 0\,, \\[0.5ex]
B^{\rm (SEVP)}_{62}(n{\rm S}) &=& \frac{8}{45} = 
0.177\,778\,, \\[0.5ex]
B_{62}(1{\rm S}) &=& \frac{104}{135} - \frac{16}{9}\,\ln 2 = 
-0.461\,891\,, \\[0.5ex]
B_{62}(n{\rm S}) &=& B_{62}(1{\rm S}) 
+ \frac{16}{9} \left( \frac34 + \frac{1}{4 n^2} -
\frac1n - \ln(n) + \Psi(n) + C\right)\,,
\end{eqnarray}
\end{subequations}
where $\Psi$ denotes the logarithmic derivative of the gamma function,
and $C = 0.577216\dots$ is Euler's constant.

The formulas for $B_{61}$ are a little more involved,
\begin{subequations}
\label{B61}
\begin{eqnarray}
B^{\rm (2LSE)}_{61}(1{\rm S}) &=& \frac{127\,069}{32\,400} 
+ \frac{875}{72}\,\zeta(2) + \frac92\, \,\zeta(2)\,\ln 2 
- \frac98 \, \zeta(3)
\nonumber\\[0.5ex]
& &  - \frac{152}{27}\,\ln 2 
+ \frac{40}{9} \, \ln^2 2 + \frac43 \, N(1{\rm S}) =
49.731\,651\,, \\[0.5ex]
B^{\rm (2LSE)}_{61}(n{\rm S}) &=& B^{\rm (2LSE)}_{61}(1{\rm S})
+ \frac43 \, \left[ N(n{\rm S}) - N(1{\rm S})\right] 
\nonumber\\[0.5ex]
& & + \! \left( \frac{80}{27} - \frac{32}{9} \, \ln 2\right) 
\left(\frac34 + \frac{1}{4 n^2} - \frac1n - \ln(n) \! + \! \Psi(n) 
\! + \! C\right), \\[0.5ex]
B^{\rm (SVPE)}_{61}(n{\rm S}) &=& 
-\frac{7}{162} + \frac{5}{72}\,\zeta(2) =
0.071\,022\,, \\[0.5ex]
B^{\rm (SEVP)}_{61}(1{\rm S}) &=& 
-\frac{401}{2025} + \frac{16}{15}\,\ln 2 =
0.541\,332\,, \\[0.5ex]
B^{\rm (SEVP)}_{61}(n{\rm S}) &=& B^{\rm (SEVP)}_{61}(1{\rm S})
- \frac{32}{45}  
\left(\frac34 + \frac{1}{4 n^2} - \frac1n \! - \! \ln(n) 
\! + \! \Psi(n) \! + \!  C\right), \\[0.5ex]
B_{61}(1{\rm S}) &=& \frac{39\,751}{10\,800}
+ \frac{110}{9}\,\zeta(2) + \frac92\, \,\zeta(2)\,\ln 2
- \frac98 \, \zeta(3) \nonumber\\[0.5ex]
& &  - \frac{616}{135}\,\ln 2
+ \frac{40}{9} \, \ln^2 2 + \frac43 \, N(1{\rm S}) =
50.344\,005\,, \\[0.5ex]
B_{61}(n{\rm S}) &=& B_{61}(1{\rm S})
+ \frac43 \, \left[ N(n{\rm S}) - N(1{\rm S})\right] 
\\[0.5ex]
& & + \left( \frac{304}{135} - \frac{32}{9} \, \ln 2\right) \,
\left(\frac34 + \frac{1}{4 n^2} - \frac1n - \ln(n) + \Psi(n) +           
C\right)\,.\nonumber
\end{eqnarray}
\end{subequations}

The results for $N(n{\rm S})$ taken from~\cite{Je2003jpa} read
\begin{eqnarray}
\label{NnS}
N(1{\rm S}) &=& 17.855\,672(1)\,, \quad
N(2{\rm S}) = 12.032\,209(1)\,, \nonumber\\[0.5ex]
N(3{\rm S}) &=& 10.449\,810(1)\,, \quad
N(4{\rm S}) = 9.722\,413(1)\,, \nonumber\\[0.5ex]
N(5{\rm S}) &=& 9.304\,114(1)\,, \quad
N(6{\rm S}) = 9.031\,832(1)\,, \nonumber\\[0.5ex]
N(7{\rm S}) &=& 8.840\,123(1)\,, \quad
N(8{\rm S}) = 8.697\,639(1)\,.
\end{eqnarray}
The slightly shifted results for $N(1{\rm S})$ also explains
a discrepancy in an intermediate step of the calculation of
radiative corrections to the muonium hyperfine 
splitting~\cite{Pa1996,NiKi1997}.

%
% Extrapolation of the $Z\alpha$--Expansion 
%
\section{Extrapolation of the $Z\alpha$--Expansion}
\label{analytic}

We start our consideration with the one-loop self-energy which is
the dominating radiative correction in hydrogenlike bound systems.
We write the (real part of the) one-loop self-energy shift
$\Delta E^{\rm (1L)}_{\rm SE}$ as
\begin{equation}
\label{ESEasF}
\Delta E^{\rm (1L)}_{\rm SE} = 
\frac{\alpha}{\pi} \, (Z \alpha)^4 \,
\frac{m}{n^3} \, F(Z\alpha)\,,
\end{equation}
where $F(Z\alpha)$ is a dimensionless quantity
which depends on the principal 
quantum number $n$, the total electron spin$+$angular momentum
$j$ and the electron orbital angular momentum $l$, and of course
on the parameter $Z\alpha$.

The semi-analytic expansion of $F(Z\alpha)$
about $Z\alpha = 0$ for P states and states 
with higher angular momenta gives rise to the expression~\cite{SaYe1990},
\begin{equation}
\label{defF}
F(Z\alpha) = A_{40} + (Z \alpha)^2 \, 
\left[A_{61} \, \ln(Z \alpha)^{-2} + A_{60}\right] + \dots
\qquad (l \geq 1)\,,
\end{equation}
where the ellipsis again denotes omitted higher-order terms.
The $A_{60}$ coefficient has proven to be by far the 
most difficult to evaluate~\cite{ErYe1965a,ErYe1965b,%,
Er1971,Sa1981,Pa1993}, and for 2P states, results have become 
available recently~\cite{JePa1996,JeSoMo1997}.

%
% Figure 2 
%
\begin{figure}[htb!]
\begin{center}
\begin{minipage}{12.0cm}
\begin{center}
\centerline{\mbox{\epsfysize=9.0cm\epsffile{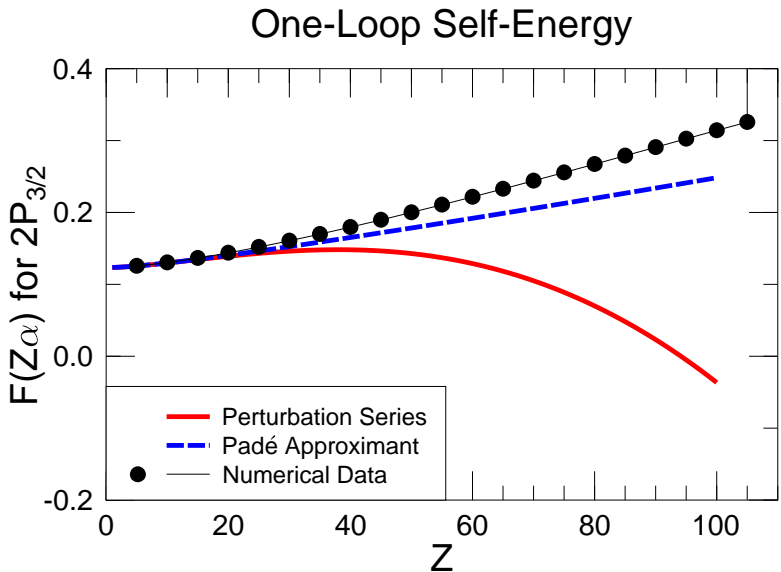}}}
\caption{\label{fig2}
Extrapolation of the semi-analytic
$Z\alpha$-expansion of the one-loop self-energy
(\ref{defF}) to the range of high
nuclear charge via ``deferred'' $[2/2]$--Pad\'{e}-approximants
for the $2{\rm P}_{3/2}$-state
as described in the text. The analytic coefficients 
$A_{40}$, $A_{61}$ and $A_{60}$ in Eq.~(\ref{defF})
are taken from Refs.~\cite{JePa1996,JeSoMo1997}.
The extrapolated semi-analytic $Z\alpha$-expansions are
closer to the numerical data for high $Z$ than the ``raw''
$Z\alpha$-expansion. Numerical data at high $Z$ are taken from
Refs.~\cite{Mo1974b,Mo1992}.}
\end{center}
\end{minipage}
\end{center}
\end{figure}
    
%
% Figure 3
%
\begin{figure}[htb!]
\begin{center}
\begin{minipage}{12.0cm}
\centerline{\mbox{\epsfysize=9.0cm\epsffile{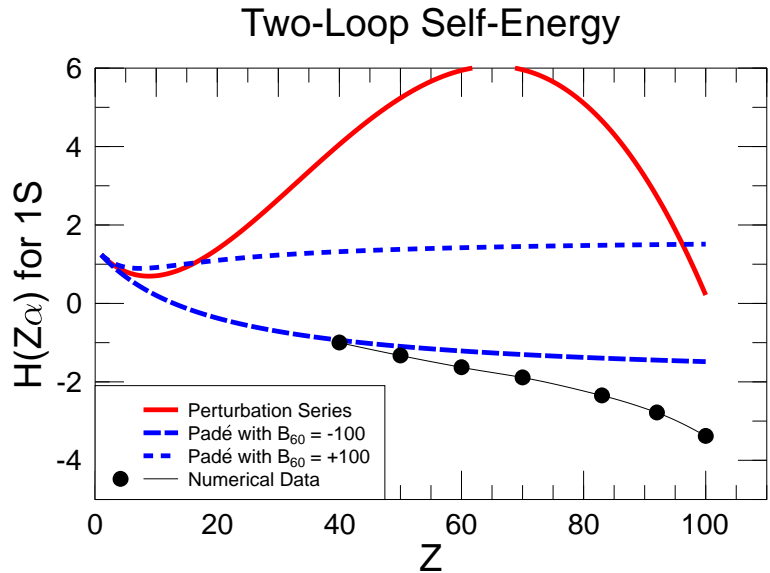}}}
\caption{\label{fig3}
Extrapolation of the two-loop self-energy.
Analytic data are taken from Sec.~\ref{review}
[see the (2LSE)-parts of Eqs.~(\ref{B40})---(\ref{NnS})]. 
Numerical data are found in Ref.~\cite{YeInSh2003}. 
Much better agreement between
numerical and analytic data is achieved for negative $B_{60}$.}
\end{minipage}
\end{center}
\end{figure}

The semi-analytic expansion (\ref{defF}) is generally assumed
to converge to the function $F(Z\alpha)$ for low $Z$, at least
in an asymptotic sense. This is confirmed by recent numerical
evaluations~\cite{JeMoSo1999,JeMoSo2001pra} for S and P states
and the successful
consistency check with available analytic results~\cite{JePa1996,Pa1993}.
In many cases, an asymptotic expansion valid {\em a priori} for small
expansion parameter $Z\alpha$ can be extrapolated to large coupling,
if it is combined with a suitable convergence acceleration
or resummation method (the latter in the case of a divergent input
series~\cite{We1989}). The logarithms in Eq.~(\ref{defF}) make 
a power series expansion about $Z\alpha=0$ impossible.
However, an extrapolation is still possible if we 
expand about $Z\alpha=\alpha\neq 0$ and use the fact that 
the nonperturbative function $F(Z\alpha)$, in the 
range of small $Z$, is very well represented
by the first terms in its asymptotic expansion, as suggested
by Fig.~1 of Ref.~\cite{JeMoSo1999}. 
The logarithms in (\ref{defF}), when expanded about
$Z\alpha=\alpha$, give rise to an infinite power series
in the variable $g = (Z-1)\,\alpha$. 
We proceed as follows: for all radiative corrections
studied in the sequel, we start from the semi-analytic
expansion and take into account all known coefficients.
We then expand in $g$ and
evaluate the diagonal $[2/2]$--Pad\'{e} approximant
to the resulting power series (for the definition and a
comprehensive discussion of
Pad\'{e} approximants we refer to~\cite{BaGr1996}).
This could be characterized a ``deferred'' approximant
which is evaluated only after one has ``advanced'' to the 
point $Z = 1$ from the ``starting point'' $Z=0$
(or equivalently $Z\alpha = 0$). Formally,
the semi-analytic $Z\alpha$-expansion is performed
about the point $Z=0$. In re-expanding the perturbation series
about a different point in the complex plane,
we follow ideas outlined in~\cite{Re1982} which were 
originally applied to the problem of calculating
the autoionization width of atomic resonances in an 
external electric field.

The choice of the $[2/2]$--Pad\'{e} approximant
is motivated by the paradigm of
finding a compromise between the necessity to harvest 
the information contained
in the logarithms, which give rise to power series terms
of arbitrarily high order,
and at the same time to avoid spurious singularities
which may be incurred when the degree of the 
Pad\'{e} approximant is increased to an excessively high order.
The $[2/2]$--deferred Pad\'{e} approximant
about $Z=1$ to the function $F$ 
[see also~Eq.~(3) of~\cite{JeBeWeSo2000}] has five parameters
$p_0,\dots,p_2$, $q_1,\dots,q_2$,
\begin{equation}
\label{def22}
[2/2]_F(g) = \frac{\sum_{i=0}^{2} p_i\,g^i}
  {1 + \sum_{j=1}^2 q_j\,g^j}\,,
\end{equation}
which are determined by the requirement that the power series expansion
of $[2/2]_F(g)$ about $g=0$ reproduce the power series expansion
of $F(Z\alpha)$ with $Z\alpha = \alpha + g$, also about $g=0$, up to the 
order ${\mathcal O}(g^4)$. As discussed in~\cite{BaGr1996}, this 
condition alone defines the Pad\'{e} approximant uniquely. 
The ``one'' in the denominator of (\ref{def22}) is the so-called
Baker convention.

Because we do not observe a factorially divergent alternating-sign
pattern in the $g$-expansion, we do not
employ the delta transformation~\cite[Ch.~8]{We1989} which
has proven to be superior to Pad\'{e} approximants in a number of
applications where factorial divergence is observed 
(e.g.,~\cite{JeBeWeSo2000}).
We rely on the robust Pad\'{e} approximants, while stressing that
it may be possible to find better extrapolation algorithms that
harvest the analytic structure of (\ref{defF}) and give
rise to logarithmic terms {\em naturally}. As yet, we have been
unable to find such algorithms.

Fig.~\ref{fig2}--\ref{fig3} show that the
extrapolated semi-analytic expansions have a somewhat 
better agreement
with medium and high--$Z$ numerical data than
the known terms of the $Z\alpha$-expansion alone. 
We observe that the high--$Z$ results for the 
energy correction given by the
irreducible set of two-loop self-energy insertions  
into the bound electron propagator (see Ref.~\cite{YeInSh2003})
can only be made consistent with our extrapolated
$Z\alpha$-expansion if we assume that the coefficient 
$B_{60}(1{\rm S}_{1/2})$ is negative and rather large in magnitude.

%
% Conclusions
%
\section{Conclusions}
\label{conclusions}

We have presented a ``mini-review''
of recent two-loop self-energy calculations~\cite{Pa2001,Je2003jpa,Ka1996}
in Sec.~\ref{review}, clarifying the distribution of 
sixth-order (in $Z\alpha$) two-loop binding corrections to the 
Lamb shift over the set of diagrams shown in Fig.~\ref{fig1}.
Results for the two-loop coefficients, including excited S states,
are provided in Eqs.~(\ref{B40})---(\ref{NnS}).
The distribution of the logarithmic corrections over distinct sets of diagrams
needs to be clarified in order to allow for an accurate comparison to 
numerical calculations which are currently being pursued~\cite{YeInSh2003}.
In Sec.~\ref{analytic}, we present a crude extrapolation scheme
for the extrapolation of the $Z\alpha$-expansion from low $Z$
to high $Z$. The scheme follows ideas outlined in~\cite{Re1982} and 
is based upon an expansion in the variable $g$ where
$g$ is defined as $Z\alpha = \alpha + g$. The ``deferred'' Pad\'{e}
approximant is then evaluated in terms of the variable $g$, i.e.~after
the coupling parameter $Z\alpha$ has acquired the value $\alpha \neq 0$,
``starting'' from $Z\alpha = 0$. This deferment circumvents
the problems introduced by the logarithms in Eqs.~(\ref{defH})
and (\ref{defF}); however, we stress here that it would be highly 
desirable to find better extrapolation algorithms that
harvest the analytic structure of (\ref{defH}) and (\ref{defF}) and give
rise to logarithmic terms {\em naturally}. Although the 
extrapolation scheme has problems (in some cases, we observe spurious
poles in the Pad\'{e} approximant at medium--$Z$ values), we have 
observed rather consistent improvement over the ``raw'' $Z\alpha$-expansion
with this scheme for a number of states and QED effects which we studied
using the ``deferred'' Pad\'{e}--extrapolation scheme. Details will
be presented elsewhere.
Based on our extrapolations of the two-loop effect 
in Fig.~\ref{fig3}, we would like to advance
the tentative estimate $B_{60}(1{\rm S}_{1/2}) \approx -100(50)$.

\section*{Acknowledgements}

The author would like to acknowledge insightful and helpful discussions
with Professor Krzysztof Pachucki. Continued stimulating and 
helpful discussions with Professor Gerhard Soff are also gratefully 
acknowledged. Helpful remarks by Vladimir Yerokhin on the $B_{61}$-coefficient
also contributed to the final version of the Letter. 

%
% References
%


\begin{thebibliography}{10}

\bibitem{MoTa2000}
P.~J. Mohr and B.~N. Taylor, Rev. Mod. Phys. {\bf 72},  351  (2000).

\bibitem{Pa2001}
K. Pachucki, Phys. Rev. A {\bf 63},  042503  (2001).

\bibitem{Je2003jpa}
U.~D. Jentschura, J. Phys. A {\bf 36},  L229  (2003).

\bibitem{YeInSh2003}
V.~A. Yerokhin, P. Indelicato, and V.~M. Shabaev, e-print 
  \href{http://arXiv.org/abs/hep-ph/0302268}{hep-ph/0302268} (2003).

\bibitem{EiGrSh2001}
M.~I. Eides, H. Grotch, and V.~A. Shelyuto, Phys. Rep. {\bf 342},  63  (2001).

\bibitem{JePa2002}
U.~D. Jentschura and K. Pachucki, J. Phys. A {\bf 35},  1927  (2002).

\bibitem{ApBr1970}
T. Appelquist and S.~J. Brodsky, Phys. Rev. A {\bf 2},  2293  (1970).

\bibitem{Pa1994prl}
K. Pachucki, Phys. Rev. Lett. {\bf 72},  3154  (1994).

\bibitem{EiSh1995}
M.~I. Eides and V.~A. Shelyuto, Phys. Rev. A {\bf 52},  954  (1995).

\bibitem{Pa1993pra}
K. Pachucki, Phys. Rev. A {\bf 48},  2609  (1993).

\bibitem{EiGrSh1997}
M.~I. Eides, H. Grotch, and V.~A. Shelyuto, Phys. Rev. A {\bf 55},  2447
  (1997).

\bibitem{Ka1996}
S.~G. Karshenboim, J. Phys. B {\bf 29},  L29  (1996).

\bibitem{Ka1996b}
S.~G. Karshenboim, JETP {\bf 82},  403  (1996), [ZhETF {\bf 109}, 752 (1996)].

\bibitem{Ka1997}
S.~G. Karshenboim, Z. Phys. D {\bf 39},  109  (1997).

\bibitem{Pa1996}
K. Pachucki, Phys. Rev. A {\bf 54},  1994  (1996).

\bibitem{NiKi1997}
M. Nio and T. Kinoshita, Phys. Rev. D {\bf 55},  7267  (1997).

\bibitem{JePa1996}
U.~D. Jentschura and K. Pachucki, Phys. Rev. A {\bf 54},  1853  (1996).

\bibitem{JeSoMo1997}
U.~D. Jentschura, G. Soff, and P.~J. Mohr, Phys. Rev. A {\bf 56},  1739
  (1997).

\bibitem{Mo1974b}
P.~J. Mohr, Ann. Phys. (N. Y.) {\bf 88},  52  (1974).

\bibitem{Mo1992}
P.~J. Mohr, Phys. Rev. A {\bf 46},  4421  (1992).

\bibitem{SaYe1990}
J. Sapirstein and D.~R. Yennie,  in {\em Quantum Electrodynamics}, edited by T.
  Kinoshita (World Scientific, Singapore, 1990), pp.\ 560--672.

\bibitem{ErYe1965a}
G.~W. Erickson and D.~R. Yennie, Ann. Phys. (N. Y.) {\bf 35},  271  (1965).

\bibitem{ErYe1965b}
G.~W. Erickson and D.~R. Yennie, Ann. Phys. (N. Y.) {\bf 35},  447  (1965).

\bibitem{Er1971}
G.~W. Erickson, Phys. Rev. Lett. {\bf 27},  780  (1971).

\bibitem{Sa1981}
J. Sapirstein, Phys. Rev. Lett. {\bf 47},  1723  (1981).

\bibitem{Pa1993}
K. Pachucki, Ann. Phys. (N. Y.) {\bf 226},  1  (1993).

\bibitem{JeMoSo1999}
U.~D. Jentschura, P.~J. Mohr, and G. Soff, Phys. Rev. Lett. {\bf 82},  53
  (1999).

\bibitem{JeMoSo2001pra}
U.~D. Jentschura, P.~J. Mohr, and G. Soff, Phys. Rev. A {\bf 64},  042512
  (2001).

\bibitem{We1989}
E.~J. Weniger, Comput. Phys. Rep. {\bf 10},  189  (1989).

\bibitem{BaGr1996}
G.~A. Baker and P. Graves-Morris, {\em Pad\'{e} Approximants}, 2nd ed.
  (Cambridge University Press, Cambridge, 1996).

\bibitem{Re1982}
W.~P. Reinhardt, Int. J. Quantum Chem. {\bf 21},  133  (1982).

\bibitem{JeBeWeSo2000}
U.~D. Jentschura, J. Becher, E.~J. Weniger, and G. Soff, Phys. Rev. Lett. {\bf
  85},  2446  (2000).

\end{thebibliography}
\end{document}